\documentclass[conference]{IEEEtran}
\IEEEoverridecommandlockouts

\usepackage{cite}
\usepackage{amsmath,amssymb,amsfonts}
\usepackage{algorithmic}
\usepackage{graphicx}
\usepackage{textcomp}
\usepackage{xcolor}
\usepackage{multirow}
\usepackage{graphicx}
\usepackage{hyperref}

\usepackage[utf8]{inputenc}
\usepackage{booktabs}
\usepackage{subfig}
\usepackage{xspace}
\usepackage{siunitx}
\usepackage[markup=underlined]{changes}
\usepackage{todonotes}

\makeatother
\def\BibTeX{{\rm B\kern-.05em{\sc i\kern-.025em b}\kern-.08em
    T\kern-.1667em\lower.7ex\hbox{E}\kern-.125emX}}

\usepackage{amssymb}% http://ctan.org/pkg/amssymb
\usepackage{pifont}% http://ctan.org/pkg/pifont
\newcommand{\pb}[1]{\vspace{0.75ex}\noindent{\bf \em #1}\hspace*{.3em}}

\DeclareMathSymbol{*}{\mathbin}{symbols}{"03} % \ast
\DeclareMathSymbol{\ast}{\mathbin}{symbols}{"03}

\newcounter{RZNumberOfComments}
\stepcounter{RZNumberOfComments}

\definechangesauthor[color=red]{Koosha}
\definechangesauthor[color=blue]{Reza}
\definechangesauthor[color=magenta]{Gareth}

\setcommentmarkup{\todo[color={authorcolor!20},size=\scriptsize]{#3: #1}}

\def\eg{\emph{e.g. }\xspace}
\def\etc{\emph{etc. }\xspace}

\newcommand{\one}{({\em i}\/)}
\newcommand{\two}{({\em ii}\/)}
\newcommand{\three}{({\em iii}\/)}
\newcommand{\four}{({\em iv}\/)}
\newcommand{\five}{({\em v}\/)}

\begin{document}

% \title{The Growth of Impersonators on Instagram: \\A Deep Neural Approach
% }

% \title{A deep neural approach for impersonator\\ content detection on social media
% }

\title{Impersonation on Social Media: A Deep Neural Approach to Identify Ingenuine Content}
% \vspace{-0.4cm}

\vspace{-0.2cm}
\author{\IEEEauthorblockN{Koosha Zarei\IEEEauthorrefmark{1}, Reza Farahbakhsh\IEEEauthorrefmark{1}, No\"{e}l Crespi\IEEEauthorrefmark{1}, Gareth Tyson\IEEEauthorrefmark{2}}
\IEEEauthorblockA{\IEEEauthorrefmark{1}\textit{Institut Polytechnique de Paris, T\'el\'ecom SudParis} \textit{Evry, France}.\\
%\IEEEauthorblockA{\IEEEauthorrefmark{3}\textit{TOTAL SA., Strategy \& Innovation, EP/SG/ISB/STI, Data Analytics Competence Center (DACC)}}
\{koosha.zarei, reza.farahbakhsh, noel.crespi\}@telecom-sudparis.eu}
\IEEEauthorrefmark{2}\textit{Queen Mary University of London, United Kingdom}. gareth.tyson@qmul.ac.uk
}
% \vspace{-0.4cm}

\maketitle

%%%%%%%%%%%%%%%%%%%%%%%%%%%%%%%%%%%%%%%%%%%%%%%%%%%%%%%%%%%%%%%%%%%%%%
\begin{abstract}
Impersonators are playing an important role in the production and propagation of the content on Online Social Networks, notably on Instagram. These entities are nefarious fake accounts that intend to disguise a legitimate account by making similar profiles and then striking social media by fake content, which makes it considerably harder to understand which posts are genuinely produced.
In this study, we focus on three important communities with legitimate verified accounts. Among them, we identify a collection of 2.2K impersonator profiles with nearly 10k generated posts, 68K comments, and 90K likes. Then, based on profile characteristics and user behaviours, we cluster them into two collections of `bot' and `fan'. 
In order to separate the impersonator-generated post from genuine content, we propose a Deep Neural Network architecture that measures `profiles' and `posts' features to predict the content type: `bot-generated', 'fan-generated', or `genuine' content.
% Finally, by using advanced NLP techniques, we study what has been published by impersonators. We discovered bots regularly target specific subjects in different communities.
Our study shed light into this interesting phenomena and provides interesting observation on bot-generated content that can help us to understand the role of impersonators in the production of fake content on Instagram.

\end{abstract}


%%%%%%%%%%%%%%%%%%%%%%%%%%%%%%%%%%%%%%%%%%%%%%%%%%%%%%%%%
\section{Introduction}
%%%%%%%%%%%%%%%%%%%%%%%%%%%%%%%%%%%%%%%%%%%%%%%%%%%%%%%%%

Impersonation is where (sometimes malicious) users create social media accounts mimicking a legitimate account \cite{kooshaDeep}.
Fr example, impersonators or imposters maybe accounts that pretend to be someone popular or a representative of a known brand, company, \etc{}
Such impersonators are found on all major social media platforms. Instagram is widely used by celebrities, influencers, businesses, and public figures with different levels of popularity. 
% The potential of social networks is often mistreated by malicious users who obtain individual information from genuine verified accounts. 
Although many impersonators may be innocuous, there also exists malicious fake accounts. These often have clear plans, where they make accounts appear more popular than they are, produce pre-planned untrustworthy content, perform brand abuse or generate fake engagement \cite{zarei2020impersonators}. 
%From malicious activities in social media, a larger set of threats have been identified including brand abuse, fraud and follower farming. 
Therefore several lawsuits have taken place in the United State (along with other countries), where criminal impersonation is a crime. It involves assuming a false identity with the intent to defraud another or pretending to be a representative of another person or organisation \cite{impersonaionlaw}. 
However, identifying such activities is often slow and laborious --- hence, developing techniques for automated detection would have real value to social media companies. In this paper, we aim to identify impersonator-generated content in Instagram. Towards that end, we pick three different and important communities with verified genuine accounts inside each. Through the pool of collected public content, by using the methodology presented in \cite{Zare1910:Typification}, we identify a set of 2.2K impersonator accounts. Next, by using unsupervised learning techniques, we find two notable clusters: \one{} `\textit{Cluster 0 - Bots}' that represent bot entities, and \two{} `\textit{Cluster 2 - Fans}' which represent fan entities. 
In this study, bots are fake accounts or social bots that tend to mimic the real user and accomplish a specific purpose \cite{grimme2017social} and interacts with humans on social media \cite{Ferrara:2016:RSB:2963119.2818717}. 
In contrast, fans are (semi-) human-operated accounts that are created and maintained by a fan or devotee about a celebrity, thing or particular phenomenon.
We then use these clusters to create necessary labels for building and training a Deep Neural Network to predict post types: \one{} bot-generated, \two{} fan-generated, or \three{} genuine content.
% We finally we focus on the published content of impersonators to shed light on their behavioral patterns. We leverage natural language processing (NLP) techniques to understand post captions,  get written topics and sentiments, and compare results to genuine content. 
The contribution of this study can be summarised as follow:

\begin{itemize}

\item We assemble a novel dataset containing the content and activities of impersonators in three leading communities: Politicians, Sport Starts and Musicians.

\item We present a practical approach to cluster impersonators and generate content labels based on profile characteristics and user behaviours. 

\item We propose a Deep Neural Network architecture in order to detect and predict impersonator-generated posts and genuine content. 

%\item We present an investigation of the generated content by impersonators and their behaviours. 
%\reza{removed for short paper of 5 pages}

\end{itemize}

\vspace{-0.2cm}
\section{Methodology \& Dataset}
\label{data_collection}
% We start by defining impersonator, its variations, and terms that are used in this study and then, we describe our data collection methodology. %%In general, we compile a set of popular case study accounts on Instagram and we identify their impersonators.

% \begin{figure}[htbp]
% \vspace{-0.4cm}
% \centerline
% {\includegraphics[width=0.45\textwidth]{Figures/impersonator_taxonomy.jpg}}
% \vspace{-0.3cm}
% \caption{The taxonomy of impersonators.}
% \label{fig_impersonator_taxonomy}
% \vspace{-0.4cm}
% \end{figure}

\subsection{Definition and Taxonomy}\label{impersonator_taxonomy}

\pb{Bots:} are (semi-) automatic agents that are designed to accomplish a specific purpose \cite{grimme2017social} and automatically produce content and interacts with humans on social media \cite{Ferrara:2016:RSB:2963119.2818717}. Bots are normally defined with the condition of mimicking human behaviour \cite{stieglitz2017social}.

\pb{Impersonator or Imposter:} is someone on social media who builds a profile using the information of another legitimate account and pretends to be that entity or copies the behaviour/actions of that profile \cite{zarei2020impersonators}. 

\begin{figure}[t!]
\vspace{-0.3cm}
\centerline
{\includegraphics[width=0.42\textwidth]{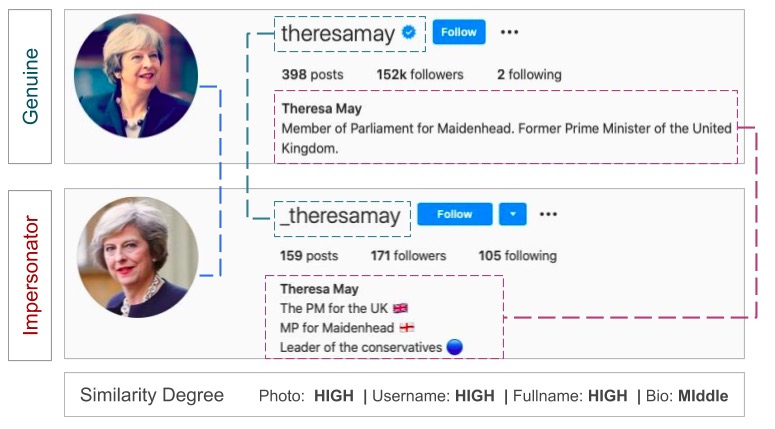}}
\vspace{-0.2cm}
\caption{Identifying Impersonators through profile similarity.}
\label{fig_impersonator_example}
\vspace{-0.5cm}
\end{figure}

\pb{Profile Similarity:} We use this term to indicate whether there is any similarity or correlation between two Instagram profiles. Similarity can be in \one{} text features \cite{10.1007/978-3-642-41278-3_74} such as username, full name, or biography \eg{} `\textit{@barackobama}' and `\textit{@barack\_\_obama}', or \two{} profile photos (if the same person exists in both photos). The `\textit{Similarity Level}' could be high (similar in all metrics),  low (just in one metric), or between. An example of the genuine Theresa May account and her impersonator with a high degree of similarity is shown in Figure \ref{fig_impersonator_example}.
In \cite{kooshaDeep} \cite{Zare1910:Typification} we introduced the problem of impersonation and discussed the identification methods. Then, we uncovered unknown groups of impersonators and examined their behaviours. For example, fan pages have a higher number of followers and are completely public pages. But bots, have very fewer followers and publish a lot of posts in a shorter period of time. Then, in \cite{zarei2020impersonators} we studied the comments they generated under the post of genuine figures in details. For example, bots produce much higher duplicated comments than others and give likes (passive reaction) faster.
Eventually, in this study, we divide impersonators into two broad types of public accounts:

% \begin{enumerate}

(1) \textbf{Bot Impersonator (Bot)}:  these public fake accounts or social bots tend to mimic the real user and generally generate specific content. 
First, from profile characteristics, bots are usually simple accounts that use default Instagram settings: no full name, no biography, and sometimes no profile photos. The follower count is low and they follow a lot of other accounts.
From similarity viewpoint (compared to a genuine user), bots have weak profile similarity degrees: they have no similar profile photo and have low similarity in username, full name, or biography.
From activity viewpoint, bots receive very limited engagement (like or comment) per post, are lazy in publishing stories, are so active in giving comments and likes to others, and the rate of issuing duplicated comments is high.
Existing bots vary in sophistication. Some bots are very simple and merely re-publish posts, whereas others are sophisticated and can even interact with human users or post comment.
In this study, `\textit{Bot Impersonator}' and '\textit{bot}' terms are interchangeable.

(2) \textbf{Fan Impersonator (Fan)}: is a (semi-) human-operated account that is created and maintained by a fan or devotee about a celebrity, thing or particular phenomenon. 
From profile perspective, fans have a greater follower number than bots, are completely public accounts, have a biography, and usually use a URL.
From impersonation viewpoint, fans have higher profile similarity in photo, username, full name, and biography metrics.
From behaviour viewpoint, fans are interested in publishing posts and stories, are more productive than bots, receive higher engagement within their posts (both like and comment), and the owner barely shares self-generated content. 
From managing viewpoint (who controls the page), we can divide fans into two different types (Figure \ref{fig_impersonator_taxonomy}):
\one{} A fan page which is regulated by `human'. In this situation, there is no automation movement and all content and activities are published by a human.
\two{} A fan page which is regulated by `human and bot'. In this type, page owner which is a human usually use some automation and bot services to gain attention. For example, using a bot to comment or like on related pages.

% \end{enumerate}

\subsection{Case Study Accounts}\label{case_studies}

To seed our analysis, we select a set of 15 ground-truth verified accounts from three communities: \textit{politicians}, \textit{sports stars}, and \textit{musicians (celebrities)}.
We pick these communities to compare the impersonation problem in divided societies. For each community, we select the top 5 most popular verified accounts manually, then we confirm the popularity by \cite{hypeauditor}:

\textbf{\textit{Politicians:}} Donald J. Trump (\textit{@realdonaldtrump}), Barack Obama (\textit{@barackobama}), Emmanuel Macron (\textit{@emmanuelmacron}), Boris Johnson (\textit{@borisjohnsonuk}), and Theresa May (\textit{@theresamay}).
\textbf{\textit{Sports Stars:}} Leo Messi (\textit{@leomessi}), Cristiano Ronaldo (\textit{@cristiano}), Rafael Nadal (\textit{@rafaelnadal}), Roger Federer (\textit{@rogerfederer}), and Novak Djokovic (\textit{@djokernole}).
\textbf{\textit{Musicians:}} Lady Gaga (\textit{@ladygaga}), Beyonce  (\textit{@beyonce}), Taylor Swift (\textit{@taylorswift}), Adele (\textit{@adele}), and Madonna (\textit{@madonna}).
% All use cases are well-known singers with verified accounts on Instagram.

\begin{table}[h!]
\vspace{-0.25cm}
\centering
\caption{Use Cases and Corresponding Hashtags}
\vspace{-0.22cm}
\label{tbl_dataset_hashtags}
\resizebox{\columnwidth}{!}{% 
% \scalebox{1}{
\begin{tabular}{ll|ll|ll}
\multicolumn{2}{c}{\textbf{Politician}} & \multicolumn{2}{c}{\textbf{Sports Stars}} & \multicolumn{2}{c}{\textbf{Musician}} \\ \hline
\multicolumn{1}{l}{D. Trump} & \textit{\#donaldtrump} & L. Messi & \textit{\#leomessi} & L. Gaga & \multicolumn{1}{l}{\textit{\#ladygaga}} \\
\multicolumn{1}{l}{B. Obama} & \textit{\#barackobama} & C. Ronaldo & \textit{\#cristianoronaldo} & Beyonce & \multicolumn{1}{l}{\textit{\#beyonce}} \\
\multicolumn{1}{l}{E. Macron} & \textit{\#emmanuelmacron} & R. Federer & \textit{\#rogerfederer} & T. Swift & \multicolumn{1}{l}{\textit{\#taylorswift}} \\
\multicolumn{1}{l}{B. Johnson} & \textit{\#borisjohnson} & R. Nadal & \textit{\#rafaelnadal} & Madonna & \multicolumn{1}{l}{\textit{\#madonna}} \\
\multicolumn{1}{l}{T. May} & \textit{\#theresamay} & N. Djokovic & \textit{\#novakdjokovic} & Adele & \multicolumn{1}{l}{\textit{\#adele}} \\ \hline
\end{tabular}
}
\vspace{-0.35cm}
\end{table}

\subsection{Data Collection} \label{s:dataset_subsection}

% Using the above set of 15 accounts, we next proceed to gather data on potential impersonators. Our data collection involves several key steps as follow: 

\pb{Genuine Accounts:} First, we collect posts of our 15 genuine case studies (listed in section
``\ref{case_studies}") which are published between October 2018 and January 2020. Posts contain publicly available information including caption, hashtags, image/video, number of likes, number of comments, location, time, and tagged list. 1.3K posts across the three communities has been collected during the campaign. We use the crawler presented in \cite{kooshaDeep}.

\pb{Identifying Impersonators:}\label{impersonator_detection}
To obtain a set of impersonators, we configure a crawler to collect public posts that contain associated hashtag with the name of each account (Table \ref{tbl_dataset_hashtags}) between September 2019 and January 2020. For example, in Trump, we gather posts include the \textit{\#donaldtrump} tag. 
Next, based on the methodology that we presented in~\cite{kooshaDeep}, we measure the profile similarity of the publishers to identify impersonators across case studies. 
The methodology is based on the Instagram profile similarity and we consider major profile metrics such as \textit{username (text), full name (text), biography (text), profile photo (image), follower count, followee count, media count, and account age}. 
\one{} For text metrics, we use the Cosine Similarity technique \cite{10.1007/978-3-642-41278-3_74} and we define the minimum threshold to 30\%. \two{} To measure the photo similarity, we use a convolutional neural network face detection in \cite{FaceRecognitionLib}. We compare the face of all accounts (if exist) to the face of the genuine users (\eg{} R. Federer) and if the same person is detected, we mark it as similar photos.
Eventually, if an account has at least 30\% similarity in one of the text metrics or has a similar profile photo, we consider it as an impersonator. Otherwise, it is a non-similar account (\emph{not} impersonator) and we exclude it from the dataset.
In total, we discover 1.6K  impersonators with different levels of similarity. 

\pb{Followers/Followees:} We next crawl the follower and followee list of each impersonator from the previous phase (October 2018 to January 2020). As it is infeasible to collect \emph{all} followers/followees, we define a limitation of 1K for followers and 500 for followees. At the same time, we examined the profile similarity of them to see if they are impersonator or not. Finally, we have 2.3K impersonators.

\pb{Posts:} We crawled the 50 most recent posts published by the impersonator. Furthermore, we gather impersonators' \one{} profile information, \two{} number of comments received on posts, and \three{} number of likes attracted on posts. This task was running simultaneously between October 2018 and January 2020.

\pb{Validation:} We finally manually inspect the profiles of the impersonators to confirm they are impersonators. We filter any incorrectly identified impersonators alongside their posts. 36 Impersonators were identified incorrectly (1.5\% of the total population), and 42 accounts  (1.8\%) change the application of the page or sell their account at some point during the measurement period.
In total, we obtain nearly 68K comments and 90K likes from 10K posts of 2.2K impersonators (Table \ref{tbl_dataset_summary}).

\begin{table}[h!]
\vspace{-0.35cm}
\centering
\caption{Summary of Dataset}
\vspace{-0.2cm}
\label{tbl_dataset_summary}
% \resizebox{\columnwidth}{!}{% 
\scalebox{1}{
\begin{tabular}{lrrrr}
\hline
\textbf{Community} & \textbf{Imposter} & \textbf{post} & \textbf{comment} & \textbf{like} \\ \hline
Politician & 36\% & 30\% & 36\% & 35\% \\
Sport player & 34\% & 30\% & 34\% & 40\% \\
Musician & 30\% & 40\% & 30\% & 25\% \\ \hline
\textbf{Total} & \textbf{2.2K} & \textbf{10K} & \textbf{68K} & \textbf{90K}
\end{tabular}
}
\vspace{-0.4cm}
\end{table}
% %ref: jupyter 01-stats.ipynb - general stat

\pb{Ethics:} In line with Instagram policies and ethical consideration on user privacy defined by the community, we only collect publicly available data through public API excluding any potentially sensitive data. 

\subsection{Data Pre-Processing}\label{text_pre_proce_sec}

\pb{Pre-Processing} Some features require pre-processing: \one{} For caption and Profile Biography, we remove all punctuation marks, stopwords and convert them to lowercase characters. We then filter words that contain fewer than three characters, and words are stemmed to reduce to their root forms. 
\two{} We then remove and covert all emojis and emoticons to word format. Then we replace URLs with `website', emails with `email', new lines with `line', and phone numbers with `phones'.
\three{} We break down each Hashtag and Username into its constituent words, \eg ``\textit{makeamericagreatagain}" contains 4 meaningful words: ``make", ``america", ``great", and ``again" \cite{wninja}. 
\four{} From posts and profile biographies, we extract hashtags (\#) and mentions (@) into separated lists.
\five{} Wherever possible, we extract the text from post image thumbnail using Tesseract OCR \cite{10.5555/1304596.1304846} and apply text pre-processing steps. 
% In Instagram, to get viewer attention, publishers sometimes prefer to put text on images/videos rather than writing a caption. 
The spaCy \cite{spacy2} is used for French Language Modeling.

%%%%%%%%%%%%%%%%%%%%%%%%%%%%%%%%%%%%%%%%%%%%%%%%%%%%%%%%%%%%%%%%%%%%%%
% \clearpage
\section{Who are Impersonators?}
\label{identification}

We start by making some primary analysis. Table \ref{tbl_imp_vs_real} presents some of the fundamental differences between real accounts and impersonators.
Impersonators tend to have few followers, but they follow many others. Normally, they do this to develop a network of relevant accounts (other impersonators) and increase their followers. Also, impersonators have a lower engagement rate.
% Furthermore, in terms of the number of comments and likes per post, impersonators suffer from lower engagement. 
% For example, with Trump, impersonators on average have 528 followers (vs. 16M), 1.1K followees (vs. 8), receive 27 comments per post (vs. 19.5K) and earn 690 likes per post (vs. 340K). We notice the same pattern for all others.

\begin{table}[t!]
% \vspace{-0.0cm}
\centering
\caption{Real Accounts vs. Impersonators}
\vspace{-0.2cm}
\label{tbl_imp_vs_real}
% \resizebox{\columnwidth}{!}{% 
\scalebox{0.75}{
\begin{tabular}{lrr|rr|rr|rr}
 & \multicolumn{2}{c}{\textbf{follower}} & \multicolumn{2}{c}{\textbf{followee}} & \multicolumn{2}{r}{\textbf{\begin{tabular}[c]{@{}r@{}}avg. \#comment \\ per post\end{tabular}}} & \multicolumn{2}{r}{\textbf{\begin{tabular}[c]{@{}r@{}}avg. \#like\\ per post\end{tabular}}} \\ \cline{2-9} 
\multirow{-2}{*}{\textit{\textbf{use case}}} & \textit{\begin{tabular}[c]{@{}r@{}}Imp\\ (avg)\end{tabular}} & \textit{\begin{tabular}[c]{@{}r@{}}real\\ account\end{tabular}} & \textit{\begin{tabular}[c]{@{}r@{}}Imp\\ (avg)\end{tabular}} & \textit{\begin{tabular}[c]{@{}r@{}}real\\ account\end{tabular}} & \textit{Imp} & \textit{\begin{tabular}[c]{@{}r@{}}real\\ account\end{tabular}} & \textit{Imp} & \textit{\begin{tabular}[c]{@{}r@{}}real\\ account\end{tabular}} \\ \hline
\multicolumn{1}{l|}{\textit{\textbf{D. Trump}}} & {\color[HTML]{680100} 528} & {\color[HTML]{036400} 16M} & {\color[HTML]{680100} 1.1K} & {\color[HTML]{036400} 8} & {\color[HTML]{680100} 27.14} & {\color[HTML]{036400} 19.5K} & {\color[HTML]{680100} 690.14} & {\color[HTML]{036400} 340K} \\
\multicolumn{1}{l|}{\textit{\textbf{B. Obama}}} & {\color[HTML]{680100} 256} & {\color[HTML]{036400} 2.5M} & {\color[HTML]{680100} 446} & {\color[HTML]{036400} 14} & {\color[HTML]{680100} 40.00} & {\color[HTML]{036400} 13.5K} & {\color[HTML]{680100} 1.4K} & {\color[HTML]{036400} 1M} \\
\multicolumn{1}{l|}{\textit{\textbf{E. Macron}}} & {\color[HTML]{680100} 435} & {\color[HTML]{036400} 1.5M} & {\color[HTML]{680100} 738} & {\color[HTML]{036400} 91} & {\color[HTML]{680100} 12.45} & {\color[HTML]{036400} 3.8K} & {\color[HTML]{680100} 302.03} & {\color[HTML]{036400} 65K} \\
\multicolumn{1}{l|}{\textit{\textbf{B. Johnson}}} & {\color[HTML]{680100} 431} & {\color[HTML]{036400} 367K} & {\color[HTML]{680100} 318} & {\color[HTML]{036400} 254} & {\color[HTML]{680100} 11.78} & {\color[HTML]{036400} 600} & {\color[HTML]{680100} 274.14} & {\color[HTML]{036400} 15K} \\
\multicolumn{1}{l|}{\textit{\textbf{T. May}}} & {\color[HTML]{680100} 312} & {\color[HTML]{036400} 157K} & {\color[HTML]{680100} 253} & {\color[HTML]{036400} 1} & {\color[HTML]{680100} 2.21} & {\color[HTML]{036400} 350} & {\color[HTML]{680100} 54.25} & {\color[HTML]{036400} 5.6K} \\ \hline
\multicolumn{1}{l|}{\textit{\textbf{Ch. Ronaldo}}} & {\color[HTML]{680100} 432} & {\color[HTML]{036400} 197M} & {\color[HTML]{680100} 832} & {\color[HTML]{036400} 445} & {\color[HTML]{680100} 12.16} & {\color[HTML]{036400} 35K} & {\color[HTML]{680100} 1.6K} & {\color[HTML]{036400} 5.5M} \\
\multicolumn{1}{l|}{\textit{\textbf{L. Messi}}} & {\color[HTML]{680100} 447} & {\color[HTML]{036400} 140M} & {\color[HTML]{680100} 650} & {\color[HTML]{036400} 227} & {\color[HTML]{680100} 13.08} & {\color[HTML]{036400} 28K} & {\color[HTML]{680100} 2.8K} & {\color[HTML]{036400} 4.1M} \\
\multicolumn{1}{l|}{\textit{\textbf{R. Nadal}}} & {\color[HTML]{680100} 121} & {\color[HTML]{036400} 8.4M} & {\color[HTML]{680100} 513} & {\color[HTML]{036400} 65} & {\color[HTML]{680100} 12.17} & {\color[HTML]{036400} 2.5K} & {\color[HTML]{680100} 768.23} & {\color[HTML]{036400} 290K} \\
\multicolumn{1}{l|}{\textit{\textbf{R. Federer}}} & {\color[HTML]{680100} 189} & {\color[HTML]{036400} 7.1M} & {\color[HTML]{680100} 479} & {\color[HTML]{036400} 71} & {\color[HTML]{680100} 9.45} & {\color[HTML]{036400} 2.9K} & {\color[HTML]{680100} 670.12} & {\color[HTML]{036400} 400K} \\
\multicolumn{1}{l|}{\textit{\textbf{N. Djokovic}}} & {\color[HTML]{680100} 148} & {\color[HTML]{036400} 6.6M} & {\color[HTML]{680100} 236} & {\color[HTML]{036400} 777} & {\color[HTML]{680100} 6.67} & {\color[HTML]{036400} 1.5K} & {\color[HTML]{680100} 320.05} & {\color[HTML]{036400} 220K} \\ \hline
\multicolumn{1}{l|}{\textit{\textbf{Lady Gaga}}} & {\color[HTML]{680100} 7.2K} & {\color[HTML]{036400} 39M} & {\color[HTML]{680100} 653} & {\color[HTML]{036400} 46} & {\color[HTML]{680100} 5.46} & {\color[HTML]{036400} 19.5K} & {\color[HTML]{680100} 219.46} & {\color[HTML]{036400} 1.1M} \\
\multicolumn{1}{l|}{\textit{\textbf{Beyonce}}} & {\color[HTML]{680100} 130} & {\color[HTML]{036400} 138M} & {\color[HTML]{680100} 701} & {\color[HTML]{036400} 0} & {\color[HTML]{680100} 3.92} & {\color[HTML]{036400} 25.8K} & {\color[HTML]{680100} 353.18} & {\color[HTML]{036400} 2.9M} \\
\multicolumn{1}{l|}{\textit{\textbf{Taylor Swift}}} & {\color[HTML]{680100} 2.4K} & {\color[HTML]{036400} 125M} & {\color[HTML]{680100} 1.3K} & {\color[HTML]{036400} 0} & {\color[HTML]{680100} 4.84} & {\color[HTML]{036400} 0.0$^{\mathrm{\textbf{*}}}$} & {\color[HTML]{680100} 177.83} & {\color[HTML]{036400} 1.8M} \\
\multicolumn{1}{l|}{\textit{\textbf{Adele}}} & {\color[HTML]{680100} 5.3K} & {\color[HTML]{036400} 33M} & {\color[HTML]{680100} 459} & {\color[HTML]{036400} 0} & {\color[HTML]{680100} 3.76} & {\color[HTML]{036400} 12.7K} & {\color[HTML]{680100} 291.15} & {\color[HTML]{036400} 1.3M} \\
\multicolumn{1}{l|}{\textit{\textbf{Madonna}}} & {\color[HTML]{680100} 6.6K} & {\color[HTML]{036400} 14.7M} & {\color[HTML]{680100} 842} & {\color[HTML]{036400} 243} & {\color[HTML]{680100} 4.74} & {\color[HTML]{036400} 1.8K} & {\color[HTML]{680100} 134.45} & {\color[HTML]{036400} 98K} \\

% \hline
\multicolumn{4}{c}
{$^{\mathrm{\textbf{*}}}$T. Swift disabled comments.}

\end{tabular}%
}
\vspace{-0.5cm}
\end{table}
% %ref: jupyter 01-stats.ipynb - general stat

\begin{figure}[htbp]
\vspace{-0.2cm}
\centerline
{\includegraphics[width=0.4\textwidth]{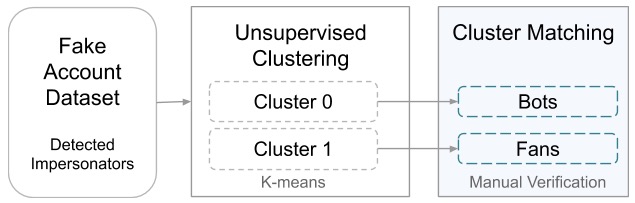}}
\vspace{-0.2cm}
\caption{The process of discovering impersonators.}
\label{fig_clustering_overview}
\vspace{-0.3cm}
\end{figure}

% \subsection{Clustering}
\pb{Clustering. }
\label{clustering}
To find the potential hidden impersonators, we do clustering. The whole process is explained in Figure \ref{fig_clustering_overview}. First, we use the impersonator dataset from section \ref{data_collection} as input and based on profile characteristics and behaviour activities, we perform unsupervised learning.
We experiment with a number of clustering methods, including K-means, Gaussian Mixture Modeling, and Spectral Clustering, finding similar results. 
Our feature list consist several features listed in Table \ref{table_clustering_features}.
This identifies two clusters  (the optimal number is obtained from the Elbow Method). The two derived clusters are highly diverse in profile characteristics and publishing behaviour (Table \ref{table_dataset_cluster_characteristics}). For the rest of this study, results are based on the K-means algorithm.
Based on manual confirmation we match discovered clusters with types of impersonators defined in section \ref{impersonator_taxonomy}. Inspection of these clusters reveals two clear populations:

\begin{table}[h!]
\vspace{-0.3cm}
\centering
\caption{Clustering Feature Set.}
\vspace{-0.2cm}
\label{table_clustering_features}
% \resizebox{\columnwidth}{!}{% 
\scalebox{1}{
\begin{tabular}{lll}
\hline
similarity username & avg received like & follower \\
similarity full name & avg hashtag length & followee \\
similarity biography & avg caption length & media count \\
similarity photo & avg received comment & private \\
external url & account age & verified \\
MSF$^{\mathrm{\textbf{*}}}$ & LSF$^{\mathrm{\textbf{*}}}$ &  \\ \hline

\multicolumn{3}{c}
{$^{\mathrm{\textbf{*}}}$The most and least number of features that have similarity.
}

\end{tabular}
}
\vspace{-0.4cm}
\end{table}
% %ref: jupyter 

\textbf{Cluster 0 - Bot}:  We believe this cluster captures bot entities (Section ``\ref{impersonator_taxonomy}") that exist to achieve specific tasks. In this study, bots are fake entities that are programmed to publish pre-defined content as posts, use a particular network of hashtags, and target specific issues. 
Bots have a quite low similarity in all profile metrics (less than 20\%) and the number of followers is almost 6 times fewer than fans (Table \ref{table_dataset_cluster_characteristics}). However, the rate of post-distribution is higher in bots. 
One of the important metrics is the received attention per post (passive or active) and bots earned nearly half of fans (almost 10 comments and  770 likes).

\textbf{Cluster 1 - Fan}:  Based on assessing characteristics, we acknowledge that this cluster represents Fans. 
Fans spread content regarding a genuine figure (in favour of or against). There is nearly 50\% similarity in the username, 40\% in the full name, 20\% in biography, and 70\% similarity in profile photos. Moreover, they hold similarity at most in 3 metrics. The number of followers is higher than the bots (avg. 101.6K vs. 16.5K) and on average, each post got 24 comments and nearly 1.6K likes (Table \ref{table_dataset_cluster_characteristics}).

% Table \ref{table_dataset_cluster_characteristics} presents a summary of the distinctive characteristics of the two clusters that have been observed.

\begin{table}[h!]
\vspace{-0.3cm}
\caption{Characteristics of the clusters.}
\begin{center}
\vspace{-0.3cm}
% \resizebox{\columnwidth}{!}{
\scalebox{0.85}{
\begin{tabular}{lrr}
\hline
\textbf{Metrics} & \textbf{Fans} & \textbf{Bots} \\ \hline
avg. username similarity per imp* & \textbf{0.49} & 0.13 \\
avg. full\_name similarity per imp & \textbf{0.40} & 0.18 \\
avg. bio similarity per imp & 0.25 & 0.18 \\
avg. photo similarity per imp & \textbf{0.71} & 0.17 \\
the Least number of features that have similarity & 1 & 1 \\
the Most number of features that have similarity & 3.32 & 1.53 \\
avg. follower per imp & \textbf{101.6K} & 16.5K  \\
avg. followee per imp & 757 & 927 \\
avg. media count per imp & \textbf{808} & 679 \\
avg. received comment per post & \textbf{24.15} & 10.01  \\
avg. received like per post & \textbf{1.6K} & 774 \\ \hline

\multicolumn{1}{c}
{$^{\mathrm{\textbf{*}}}$Impersonator}

\end{tabular}
}
\end{center}
\label{table_dataset_cluster_characteristics}
\vspace{-0.5cm}
\end{table}

\pb{Manual inspection for validation.}
\label{validation}
To validate the correctness of the proposed clustering, from each cluster we pick 80\% of profiles and check each one manually. Based on the definitions (Section \ref{impersonator_taxonomy}), 112 accounts were identified incorrectly. As we were not sure if those accounts represent a bot character or a fan entity, we recognized them as outliers and excluded from the clusters. The rest of this study is based on these validated impersonators.

% \begin{figure}[h!]
% \vspace{-0.3cm}
% \centerline
% {\includegraphics[width=0.45\textwidth]{Figures/general_overview.jpg}}
% \vspace{-0.2cm}
% \caption{The general overview of the predictor.} 
% \label{fig_general_overview}
% \vspace{-0.2cm}
% \end{figure}

%%%%%%%%%%%%%%%%%%%%%%%%%%%%%%%%%%%%%%
%%%%%%%%%%%%%%%%%%%%%%%%%%%%%%%%%%%%%%
% \clearpage

\begin{figure*}[t!]
\vspace{-0.2cm}
\centerline
{\includegraphics[width=0.7\textwidth]{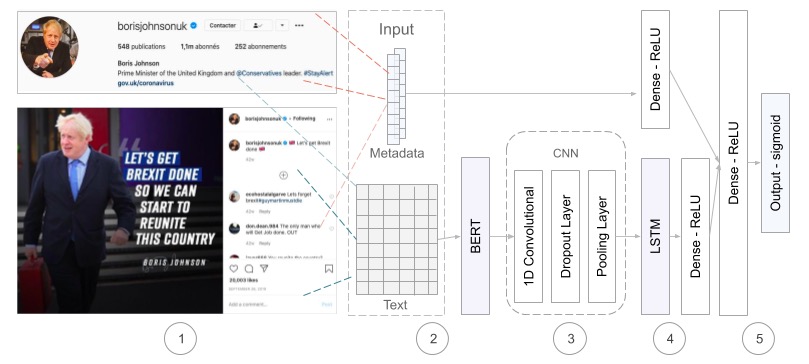}}
\vspace{-0.28cm}
\caption{The proposed Deep Neural Network architecture to detect impersonator content.}
\label{fig_dnn_architecture}
\vspace{-0.5cm}
\end{figure*}

\section{Identifying Impersonator Content}\label{classifier}

We next exploit the above dataset to explore the possibility of automatically identifying impersonator posts. 
We believe that a bot, as a fake identity, also produces untrustworthy content and fake engagements. Likewise, fan pages, in some cases may distribute fake content \eg{}a political fan page may publish rumours. 
%In terms of publishing, fans show diverse behaviours. For example, a fan page can re-share news from other pages without modifying the main content, or sometimes can add or remove a piece of information to make it more interesting, or can publish completely a rumour or fake content. Ultimately, fans can publish both correct information and also fake content.
%To answer our question, following the previous section \ref{identification} that we identify impersonators, we believe it is also possible to recognize their produced content. 
So, we use the labelled data from the previous section and present a DNN classifier to distinguish content types. This classifier can predict whether a post is impersonator-generated (fan or bot) or genuine-generated. 
Note that we do not consider the question of classifying the veracity of information shared by the accounts.

\subsection{Data Preparation}
%Here, \one{} based on features that best describe the distinctions between the bot, fan, and genuine post, \two{} the output of the clustering in section \ref{identification}, and \three{} the data that is accordingly labelled, 
% We start by presenting the design of a Deep Neural Network classifier that can detect whether a post is Impersonator-generated or not.
% The entire process is shown in Figure \ref{fig_general_overview}. 
%First, we discover impersonators in our raw dataset, and by the help clustering we uncover two fake identities among them: bots, and fans (section \ref{identification}). Then, we label posts that are generated by each cluster (genuine, bot, fan) and we verify the correctness of labelling by manual verification. Then, the output can be passed to the classifier input layer.

\pb{Dataset Overview.} For classification, we use the post dataset obtained after clustering which is described in Section \ref{data_collection}. This dataset consists of 10K post from 2.2K impersonators across 3 communities. 
Since we conduct manual annotation of impersonators, we are confident that posts are labelled correctly (pre-processing steps are discussed in Section \ref{s:dataset_subsection}).

\pb{Over-Sampling. } 
Our dataset is highly unbalanced: 
31\% genuine, 45\% fan-generated, and 34\% bot-generated post content.
To solve this problem, we use the combination of Synthetic Minority Over-sampling Technique (SMOTE) \cite{Chawla_2002} and Random Under-sampling algorithm \cite{JMLR:v18:16-365}. So, we produce similar examples from the minor class to increase the total number and, meanwhile, we under-sample the major class and randomly remove some samples. The final dataset contains an equal amount of samples from class types.
This helps us to increase the final accuracy by 8.5\%.

\begin{table}[h!]
\vspace{-0.2cm}
\centering
\caption{Feature Set used in Deep Neural Network. }
\vspace{-0.2cm}
\label{table_feature_engineering}
% \resizebox{\columnwidth}{!}{% 
\scalebox{0.75}{
\begin{tabular}{llll}
\hline
\multicolumn{2}{c}{\textbf{Post Features}} & \multicolumn{2}{c}{\textbf{Publisher Features}} \\ \hline
\textbf{Feature} & \multicolumn{1}{l|}{\textbf{Type}} & \textbf{Feature} & \textbf{Type} \\ \hline
caption text & \multicolumn{1}{l|}{\textit{text}} & similarity username & \textit{numeric} \\
caption topics (LDA) & \multicolumn{1}{l|}{\textit{text}} & similarity fullname & \textit{numeric} \\
post hashtag & \multicolumn{1}{l|}{\textit{text}} & similarity bio & \textit{numeric} \\
tagged users in post & \multicolumn{1}{l|}{\textit{text}} & profile biography & \textit{text} \\
like count & \multicolumn{1}{l|}{\textit{numeric}} & similarity photo & \textit{numeric} \\
comment count & \multicolumn{1}{l|}{\textit{numeric}} & follower/followee/post & \textit{numeric} \\
tagged users count & \multicolumn{1}{l|}{\textit{numeric}} & full name & \textit{text} \\
mention users count & \multicolumn{1}{l|}{numeric} & biography & \textit{text} \\
hashtag count & \multicolumn{1}{l|}{numeric} & username & \textit{text} \\
overall sentiment of caption & \multicolumn{1}{l|}{numeric} & following followers ratio \cite{Yang_2020} & \textit{numeric} \\
overall sentiment of hashtag & \multicolumn{1}{l|}{numeric} & followers posts ratio & \textit{numeric} \\
media type (image or video) & \multicolumn{1}{l|}{\textit{numeric}} & bio emoji count & \textit{numeric} \\
emoji count & \multicolumn{1}{l|}{\textit{numeric}} & bio hashtag count & \textit{numeric} \\
url/website exist & \multicolumn{1}{l|}{\textit{numeric}} &  & \textit{numeric} \\
date & \multicolumn{1}{l|}{\textit{numeric}} &  & \textit{} \\ \hline
\end{tabular}
}
\vspace{-0.4cm}
\end{table}
% %ref: jupyter 

\pb{Feature Engineering.} We build a set of features from post metadata and profile metrics that help us to train the proper model (Table \ref{table_feature_engineering}). We break the feature list into two principal categories:
``\textit{post features}" which comprises all features that are obtained from the content of the post such as number of likes, the caption, \etc{} And ``\textit{publisher features}" that are extracted from the profile of the publisher profile.
To prepare the feature set, we directly use some features such as numbers. However, some others are derived from the content. For example, the account age is taken from the date of the first post and the profile similarities are calculated previously in section \ref{impersonator_detection}. 
Then, to do text vectorization, the caption text, user biography, and other text metrics are vectorized using Keras Tokenizer \cite{chollet2015keras} class with 30000 num\_words. This class allows vectorizing a text corpus, by turning each text into either a sequence of integers.

\pb{Proposed DNN Architecture.} 
Then, we propose a Deep Neural Network architecture that exploits CNN, LSTM, BERT and Dence Layers to process post content and profile metadata (Figure \ref{fig_dnn_architecture}). The workflow is as follows:

(1) First, in the input layer, we extract and pre-process all features that are listed in Table \ref{table_feature_engineering}. This architecture accepts two inputs types: \one{} text content (\eg{} post caption, hashtags, profile bio) which we combine them into a single corpus. \two{} the metadata features (\eg{} like, comment, follower, followee) that come from both profile and post content and then are transformed into a single vector.

(2) Next, to transform the text into a form amenable for processing, we adopt a pre-trained language model, Bidirectional Encoder Representations from Transformers (BERT) \cite{devlin2018bert}. This, results in an output vector by BERT (vectorized text) and then given as input to a CNN layer.

(3) Then, the tokenized output of the BERT layer passes through a Convolution Neural Networks. This network contains 1D CNN with ReLU activation function (and 128 filters and a kernel size of 6) followed by a Dropout Layer (value of 0.2) for regularization, then a 1D Pooling Layer. 

(4) Then, \one{} the result of CNN layer connects to a LSTM layer which processes vectorized text data and outputs a single 32-dimensions vector that is then fed forward through a ReLU activated Dense layer of size 16. \two{} Meanwhile, numerical metadata passes through a Dense Layer with ReLU activation of size 16.

(5) Finally, we concatenate the output of the text and metadata layers into a single vector (size 32) that is then fed forward through a Dense layer with ReLU activation function and then an Output Layer which forms the type of the post (bot, fan, genuine). We develop this model using Tensorflow and Keras Functional API \cite{chollet2015keras}.

We pick a random split of 75\% (training set) and 25\% (test set) and run with 10-Fold Cross-Validation. The Accuracy, Precision, Recall, and F1-Score results are listed in Table \ref{table_model_performance}. We compare the proposed classifier with a tradition Random Forest Classifier.
The traditional RF Classifiers give approximately 77\% in all metrics (text tokenized using TF-IDF).
First, we do classification using the proposed DNN architecture with only `post content' (CNN + LSTM), and we observe an increase in overall result by nearly 2\% (Accuracy 78\%). 
Then we re-run the classifier with both `\textit{post content}' and `\textit{profile metadata}' (CNN + LSTM). This helps to improve by almost 4.5\% (Accuracy 83\%). 
Finally, we add the BERT layer to our architecture (BERT + CNN + LSTM). This step additionally assists us to improve the overall efficiency by almost 4\%, and we achieve the accuracy of 86\% in detecting post type.

\begin{table}[h!]
\vspace{-0.25cm}
\centering
\caption{Performance of the proposed architecture}
\vspace{-0.2cm}
\label{table_model_performance}
% \resizebox{\columnwidth}{!}{% 
\scalebox{0.8}{
\begin{tabular}{lrrrr}
\hline
\textbf{Model} & \textbf{Accuracy} & \textbf{Precision} & \textbf{Recall} & \textbf{F1} \\ \hline
Random Forest Classifier & 0.76 & 0.78 & 0.77 & 0.76 \\
Proposed DNN (post) & 0.78 & 0.79 & 0.76 & 0.78 \\
Proposed DNN (post + profile) & 0.83 & 0.82 & 0.83 & 0.82 \\
Proposed DNN (post + profile) + BERT & \textbf{0.86} & \textbf{0.85} & \textbf{0.86} & \textbf{0.85} \\ \hline
\end{tabular}
}
\vspace{-0.45cm}
\end{table}
\section{Conclusion}\label{future}
% \section{Conclusion and Future Work}\label{future}
This study focuses on impersonators problem and the challenge of identifying the impersonator-generated content on Instagram. First, by the help of clustering we recognised two clusters and based on their  characteristics, we clustered them as Fans and Bots. Then, in order to detect what do they publish, we introduced a DNN which can correctly classify posts as `bot-generated', `fan-generated', or `genuine' content.
% Then, we investigated the content produced by impersonators. Based on advanced NLP techniques, we observed discrete characteristics in publishing from bots and fans.
The results of this study help community on better understanding the phenomena of bot-generated content in social media.

%%%%\reza{we removed for short paper 5 pages}
%%\pb{As future work,} We believe the accuracy of the DNN architecture can be improved by looking into comments, stories, Reels as well as leverage the connection between impersonators to see the impact of bots on the propagation of fabricated content. From another perspective, it is priceless to understand what percentage of the impersonator-generated content can be considered as a type of fake information.

%%%%%%%%%%%%%%%%%%%%%%%%%%%%%%%%%%%%%%%%%%%%%%%%%%%%%%%%%%%%%%%%%%%%%%

\bibliographystyle{unsrt}
%\vspace{-0.1cm}
\bibliography{ref}

\end{document}